\documentclass[12pt]{article}

\input epsf.tex
\newcommand{\be}{\begin{equation}}
\newcommand{\ee}{\end{equation}}
\newcommand{\bea}{\begin{eqnarray}}
\newcommand{\eea}{\end{eqnarray}}

\begin{document}

\begin{titlepage}

\begin{flushright}
UUITP-06/01\\ 
hep-th/0110265
\end{flushright}

\vspace{1cm}

\begin{center}
{\Large\bf A black hole hologram\\
\smallskip

in de Sitter space \\}

\end{center}
\vspace{3mm}

\begin{center}

{\large Ulf H.\ Danielsson

\vspace{5mm}

Institutionen f\"or teoretisk fysik  \\
Box 803, SE-751 08
Uppsala, Sweden}

\vspace{3mm}

{\tt
ulf@teorfys.uu.se\\}

\end{center}

\vspace{5mm}

\begin{center}
{\large \bf Abstract}
\end{center}
\noindent

In this paper we show that the entropy of de Sitter space with a black hole in arbitrary dimension can be
understood using a modified Cardy-Verlinde entropy formula. We also comment on the observer dependence
of the de Sitter entropy.

\vfill
\begin{flushleft}
October 2001
\end{flushleft}
\end{titlepage}
\newpage


\section{Introduction}

\bigskip

Physics is very different depending on the presence and sign of the
cosmological constant. With a vanishing cosmological constant, space time is
asymptotically flat and physics can be conveniently described by an
S-matrix. This is a consequence of the presence of a light like infinity; if
we wait long enough particles may be separated by an arbitrary spatial
distance and interactions will be suppressed. In the case of a negative
cosmological constant the universe is anti de Sitter (AdS) on large scales
and there is no light like infinity. There is, however, at an infinite
spatial distance a time like infinity which can be used as a holographic
screen of Lorentzian signature, \cite{Maldacena:1998re}, \cite{Gubser:1998bc}%
, \cite{Witten:1998qj}. This discovery was the first rigorous implementation
of the idea of holography, \cite{'tHooft:1993gx}, \cite{Susskind:1995vu},
and has lead to many subsequent studies.

But it is very likely that our universe is neither flat nor AdS but instead
is a de Sitter space (dS) on large scales with a positive cosmological
constant, \cite{Riess:1998cb}, \cite{Perlmutter:1999np}. While being the
most realistic possibility, de Sitter space is at the same time
theoretically the most challenging. To this date there is, in fact, no
successful implementation of de Sitter space in string theory.

One of the intriguing features of de Sitter space is the presence of a
horizon and an associated temperature and entropy, \cite{Gibbons:1977mu}.
While string theory successfully has addressed the problem of entropy for
black holes, dS entropy remains a mystery. One reason is that the finite
entropy seems to suggest that the Hilbert space of quantum gravity for
asymptotically de Sitter space is finite dimensional, \cite{Banks:2000fe}. A
recent discussion of this and other connected issues can be found in \cite
{Witten:2001kn}. Another, related, reason is that the horizon and entropy in
de Sitter space have an obvious observer dependence. For a black hole in
flat space (or even in AdS) we can take the point of view of an outside
observer who can assign a unique entropy to the black hole. The problem of \
what an observer venturing inside the black hole experiences, is much more
tricky and has not been given a satisfactory answer within string theory.
While the idea of black hole complementarity provides useful clues, \cite
{Susskind}, rigorous calculations are still limited to the perspective of
the outside observer. In de Sitter space there is no way to escape the
problem of the observer dependent entropy. This contributes to the
difficulty of de Sitter space.

Recently there has been some progress towards a holographic understanding of
de Sitter space, \cite{Witten:2001kn}, \cite{Strominger:2001pn}. See also 
\cite{Hull:1998vg}\cite{Balasubramanian:2001rb}\cite{Hull:2001ii}. A review
with references may be found in \cite{Spradlin:2001pw}. The main observation
is that there is a possibility of introducing a holographic screen at either
time like past infinity, $\mathcal{I}^{-}$, or time like future infinity, $%
\mathcal{I}^{+}$. The theory on the screen will, just as in the case of AdS,
be a conformal field theory with a scale that encodes the dimension
transverse to the screen. But in contrast to the AdS case the holographic
theory of de Sitter space will be Euclidean and it is time that is given a
description through the scale. Large scales on $\mathcal{I}^{+}$ will
correspond to early times, while late times will correspond to small scales.
The central charge of the theory is given by the area of the horizon. Very
recently it was suggested, \cite{Strominger:2001gp}, (see also \cite
{Balasubramanian:2001nb}) that our universe is described by a RG flow in the
Euclidean theory on $\mathcal{I}^{+}$. In the IR the theory has a fixed
point of relatively low central charge corresponding to an early phase of
inflation with a horizon a few orders of magnitude larger than the Planck
scale. In the UV, on the other hand, there is a fixed point of large central
charge corresponding to the universe we now are approaching where a
cosmological constant again will be dominating.

In this paper we will continue the study of a holographic description of de
Sitter space. In particular we will investigate the properties of black hole
holograms in de Sitter space. We will show that the entropy of the
cosmological horizon in the presence of a black hole can be understood using
a Cardy-Verlinde formula, \cite{Verlinde:2000wg}, in a way very similar to
the entropy of black holes in AdS. We will also make some speculations on
how to address the problem of observer dependence.

\section{Constructing the hologram}

\bigskip

\subsection{The metric}

To construct the hologram we must find appropriate coordinates to use when
we approach the space like boundary. We will concentrate on $\mathcal{I}^{+}$
which seems to be the most reasonable choice in a realistic cosmology. While
our universe is expected to become exactly de Sitter in the future, there
may very well have been a messy beginning before the inflationary de Sitter
space in the early universe which precludes the existence of a well defined $%
\mathcal{I}^{-}$. Clearly there are many possible coordinate systems to
choose among, but we will concentrate on two important and interesting
possibilities: static and cosmological coordinates respectively. Both can be
used to approach $\mathcal{I}^{+}$ and will lead to different coordinates
for the boundary theory. The static choice will lead to what we will denote
as cylindrical coordinates, while the cosmological choice will lead to
planar coordinates. We will now consider these two possibilities in turn.

\bigskip

\subsubsection{Cylindrical coordinates}

\bigskip

The d-dimensional de Sitter black hole in static coordinates is given by 
\begin{equation}
ds^{2}=-Vdt^{2}+V^{-1}dr^{2}+r^{2}d\Omega _{d-2}^{2},  \label{cylmetric}
\end{equation}
where $V=1-\frac{r^{2}}{R^{2}}-\frac{2M}{r^{d-3}}$. Throughout the paper we
will put $G=1$ where $G$ is Newtons constant. The advantage of these
coordinates is their obvious simplicity and time independence. The
disadvantage is that the expansion of the universe is not manifest. As is
clearly seen there are in general two kinds of horizons, one associated with
the black hole and one cosmological. $\mathcal{I}^{+},$which is in the focus
of our interest, is located outside the future cosmological horizon where $r$
is time like and $t$ space like. For $r\gg R$ the metric becomes 
\[
ds^{2}=-\frac{R^{2}}{r^{2}}dr^{2}+\frac{r^{2}}{R^{2}}\left(
dt^{2}+R^{2}d\Omega _{d-2}^{2}\right) . 
\]
It is clear that $\mathcal{I}^{+}$ is approached for large $r$ and we note
that the boundary is mapped on to an Euclidean cylinder $\mathbf{R}\times
S^{d-2}$ where $t$ is the coordinate along the cylinder. This kind of
coordinates were used in \cite{Balasubramanian:2001nb}.

\subsubsection{Planar coordinates}

\bigskip

Another convenient coordinate system is obtained by introducing 
\begin{equation}
\left\{ 
\begin{array}{c}
\rho =\rho \left( t,r\right) =re^{-\frac{\tau }{R}} \\ 
\tau =\tau \left( t,r\right) =t+\frac{R}{2}\ln \left( \frac{r^{2}}{R^{2}}%
-1\right) .
\end{array}
\right.  \label{plantocyl}
\end{equation}
With no black hole this becomes an inflating cosmology according to 
\[
-d\tau ^{2}+e^{2\tau /R}\left( d\rho ^{2}+\rho ^{2}d\Omega _{d-2}^{2}\right)
, 
\]
i.e. the cosmological form of the de Sitter space. These coordinates have
the obvious advantage of making the expansion of the universe explicit with
a Hubble constant given by $1/R$. Without a black hole the geodesics are
simply given by constant comoving spatial coordinates. One should note,
however, that this metric only covers half of the space time. (Another half,
bounded by\ $\mathcal{I}^{-}$, is covered by changing $e^{\frac{2\tau }{R}}$
to $e^{-\frac{2\tau }{R}}$.) $\mathcal{I}^{+}$ is now approached for large $%
\tau $ and it is clear that it will be described by spherical coordinates on
the plane. Applying the same coordinate change to the metric with a black
hole leads to the slightly more complicated metric 
\begin{equation}
ds^{2}=-N_{\tau }^{2}d\tau ^{2}+h\left( d\rho +N_{\Sigma }d\tau \right)
^{2}+r^{2}d\Omega _{d-2}^{2}  \label{planmetric}
\end{equation}
where, with $V_{0}=1-\frac{r^{2}}{R^{2}}$, we have 
\[
h=\frac{r^{2}}{V}\left( 1-\frac{r^{2}V^{2}}{R^{2}V_{0}^{2}}\right) \rho
^{-2},\qquad N_{\Sigma }=\frac{1-\frac{V^{2}}{V_{0}^{2}}}{1-\frac{r^{2}V^{2}%
}{R^{2}V_{0}^{2}}}\frac{\rho }{R}\qquad \mathrm{and}\qquad N_{\tau }=\frac{%
\sqrt{V}}{\sqrt{1-\frac{r^{2}V^{2}}{R^{2}V_{0}^{2}}}}. 
\]

If we fix $r$ and make a translation in $t$ we see from (\ref{plantocyl})
that this corresponds to a scaling of $\rho $ that leaves the metric
invariant even in the presence of the black hole. The crucial property that
makes this possible is that the metric can be written in the static, time
independent form above. Note however, that the corresponding Killing vector
is not globally time like.

\subsection{The Brown-York tensor}

We have now obtained coordinates which can be used to describe the
holographic theory. We have also seen the relation between conformal
transformations in the hologram and time translations in the de Sitter
space. To proceed we must find out more about the generators of the
transformations and their corresponding conserved quantities. To this end,
let us follow the recipe of \cite{Balasubramanian:2001nb} \cite
{Balasubramanian:1999re}and calculate the Brown-York tensor, \cite
{Brown:1993br}, of the boundary which is the generator of coordinate changes
in the bulk. The subtracted Brown-York tensor is given by 
\[
T_{ij}=-\frac{1}{8\pi }\left( K_{ij}-Kh_{ij}-\frac{d-2}{R}h_{ij}-\frac{R}{d-3%
}G_{ij}+...\right) ,
\]
where the last term is lacking in $d=3$, and the ... refers to terms needed
when $d>5$. The terms added to the $K_{ij}-Kh_{ij}$ are counter terms in the
boundary theory needed to render $T_{ij}$ finite. We have that 
\[
K_{ij}=-\frac{1}{2}\left( h_{il}g^{l\mu }\nabla _{\mu }u_{j}+h_{jl}g^{l\mu
}\nabla _{\mu }u_{i}\right) 
\]
where $u^{\mu }$ is a forward pointing, time like, unit normal to surfaces
of constant $\tau $. $g_{\mu \nu }$ is the metric of the full space time
while $h_{ij}$ is the piece induced on the boundary. The conserved quantity
associated with the bulk Killing\ vector is given by 
\[
Q=\oint_{\Sigma }d^{d-2}\Omega \sqrt{h}n^{i}\xi ^{j}T_{ij},
\]
where $n^{i}$ is an outward pointing unit normal to surfaces of constant $%
\rho $ and $\xi ^{j}$ is the Killing vector. Note that $\xi ^{j}$ is space
like near the boundary. As argued in \cite{Balasubramanian:2001nb} one can
choose coordinates in such a way that $\xi ^{j}$ is proportional to $n^{i}$
with the constant of proportionality given by the appropriate lapse function.%
\footnote{%
Note that this choice is, in general, not possible for other conserved
quantities like angular momentum.}

Let us now calculate the Brown-York tensor in the two coordinate systems
that we have chosen.

\subsubsection{Planar coordinates}

The calculation in planar coordinates involves the non diagonal metric (\ref
{planmetric}). A straight forward calculation gives, in particular, 
\[
T_{\rho \rho }=-\frac{1}{8\pi }\left( d-2\right) \frac{M}{r^{d-3}}\frac{1}{%
\rho ^{2}} 
\]
at large $\tau $. In this case the $G_{ij}+...$ counter terms vanish as $%
\tau \rightarrow \infty $ and we expect the above result to be true for
arbitrary dimension. To obtain the conserved charge we use $\xi ^{j}=\frac{r%
}{\sqrt{h}}n^{i}$, where the coefficient is the proper lapse function to
make sure that our conserved quantity is conjugate to translations in $t$.
After some calculations we find that 
\[
Q=-\frac{d-2}{8\pi }\Omega _{d-2}M, 
\]
where $\Omega _{d-2}=\frac{2\pi ^{\frac{d-1}{2}}}{\Gamma \left( \frac{d-1}{2}%
\right) }$ is the surface area of the unit $d-1$ sphere.

\bigskip But how do we make contact with the CFT? From (\ref{plantocyl}) it
follows (for fixed $r$) that 
\[
\rho \frac{d}{d\rho }=-R\frac{d}{dt}, 
\]
and we therefore conclude that the energy $E$ of the CFT, conjugate to $\rho 
$, is given by 
\begin{equation}
E=-\frac{d-2}{8\pi }\Omega _{d-2}\frac{MR}{\rho },  \label{eplan}
\end{equation}
in planar coordinates. The sign follows provided that the conformal
generator acts in the direction of decreasing $\rho $, (i.e. increasing $t$%
). Let us now proceed to perform the corresponding calculation in
cylindrical coordinates.

\subsubsection{Cylindrical coordinates}

In this case the metric is a lot simpler than in the planar case, and is
just given by (\ref{cylmetric}). The result of the calculation is of the
form 
\[
E=E_{C,A}-\frac{d-2}{8\pi }\Omega _{d-2}M, 
\]
where $E_{C,A}$ corresponds to an anomalous Casimir contribution. Note that
the energy $E$ is a dimensionful quantity which transforms in a non-trivial
way when we go between the cylinder and the plane, even if we forget about
the anomaly. In cylindrical coordinates it is independent of the direction $%
t $ along the cylinder, while on the plane it has the radial dependence
given by equation (\ref{eplan}). $E_{C,A}$ vanishes for even $d$, while we
have 
\begin{eqnarray*}
E_{C,A} &=&\frac{1}{8}\qquad \mathrm{for\qquad }d=3 \\
E_{C,A} &=&\frac{3\pi R^{2}}{32}\qquad \mathrm{for\qquad }d=5.
\end{eqnarray*}
The appropriate expressions for higher dimensions need the precise form of
the higher order counter terms. Similar formulae can also be obtained in the
case of AdS space. The only differences are that the $M$ dependent term
comes with a positive sign in all dimensions, while the anomaly is negative
in $d=3$.\footnote{%
The signs are quite subtle in the case of de Sitter. Our signs agree with
those of \cite{Balasubramanian:2001nb} but are different from those of \cite
{Klemm:2001ea} which focused on $\mathcal{I}^{-}$. For more comments on
this, see the last footnote of the paper.}

\section{Interpreting the hologram}

We have now completed the first steps towards constructing the black hole
hologram. In order to investigate it further we need an additional tool in
the form of the Cardy-Verlinde entropy formula, \cite{Verlinde:2000wg}. We
will begin by recalling how the formula is applied to the case of AdS before
we proceed with the generalization to de Sitter.

\subsection{Black holes in AdS}

In \cite{Verlinde:2000wg} it was shown that a strongly coupled CFT (with an
AdS-dual) at temperature $T$ on a cylinder $\mathbf{R}\times S^{d-2}$ (where 
$S^{d-2}$ has constant radius $R$) with a central charge $c$,\footnote{%
Following \cite{Verlinde:2000wg} we define the central charge in terms of
the subextensive Casimir energy (including possible temperature dependent
terms). Since the Casimir energy is proportional to the number of massless
fields in the CFT, the same will be true for the central charge.} has an
energy given by 
\begin{equation}
E=\frac{\left( d-2\right) c}{48\pi }\frac{V}{L^{d-1}}\left( 1+\frac{L^{2}}{%
R^{2}}\right) \equiv E_{E}+E_{C}.  \label{ECV}
\end{equation}
where $V=\Omega _{d-2}R^{d-2}$ is the volume of the $S^{d-2}$.\footnote{%
Another possibility is to choose the radius of the cylinder to be the
horizon radius $r_{s}$ rather than $R$. This modifies the subsequent
calculations but leads to the same result.} The energy is the sum of an
extensive contribution $E_{E}$ and an intensive Casimir contribution $%
E_{C}>0 $.\footnote{%
Our convention for $E_{C}$ differs by a factor of 2 from \cite
{Verlinde:2000wg}.} $L$ is a parameter related to the temperature given by 
\[
T=\frac{1}{4\pi L}\left( d-1+\left( d-3\right) \frac{L^{2}}{R^{2}}\right) , 
\]
and the entropy, finally, is given by 
\[
S=\frac{c}{12}\frac{V}{L^{d-2}}=\frac{4\pi R}{d-2}\sqrt{E_{C}E_{E}}. 
\]
The form was argued on general grounds, while the detailed coefficients were
deduced from the AdS/CFT correspondence applied to black holes in AdS. We
will verify this in detail below after generalizing the expressions to de
Sitter.

Let us now investigate what is going on a little bit more carefully.
Depending on the coordinates we use, the energy of the state we are
considering will be different. If we use planar coordinates on the boundary%
\footnote{%
This corresponds, basically, to Poincare coordinates in the bulk. For more
details on the choice of vacua in the case of AdS$_{3}$ see \cite
{Danielsson:1999wt}.}, we find that the AdS space has vanishing energy,
while a black hole in AdS will have positive energy. It is useful to write
this energy as 
\[
E=\frac{E_{plan}R}{\rho } 
\]
where $E_{plan}$ is a rescaled energy given by 
\[
E_{plan}=\frac{d-2}{8\pi }\Omega _{d-2}M. 
\]
If we instead use cylindrical coordinates, we find another vacuum where the
AdS space has a shifted energy due to the anomaly (if d is odd), and we have
that 
\[
E_{cyl}=\frac{d-2}{8\pi }\Omega _{d-2}M+E_{C,A}, 
\]
where $E_{C,A}$ is the anomalous contribution to the Casimir energy. (Note
that $E_{C,A}$ is negative for $d=3$ and positive for $d=5.$) The analysis
of \cite{Verlinde:2000wg} was performed in cylindrical coordinates based on
the work of \cite{Hawking:1982dh}\cite{Witten:1998zw}. There the expressions
for e.g. the action in the presence of a black hole had been found to be
divergent and was rendered finite by subtracting the equally divergent
contribution from empty AdS. It is therefore the energy in excess of the
vacuum associated with empty AdS space (with an extensive contribution as
well as a Casimir contribution) that is given by equation (\ref{ECV}). Hence
we conclude that 
\[
E_{E}+E_{C}=\frac{d-2}{8\pi }\Omega _{d-2}M, 
\]
and the entropy is given by 
\[
S=\frac{4\pi R}{d-2}\sqrt{E_{C}\left( E_{plan}-E_{C}\right) }. 
\]
This is the generalization of the Cardy formula proposed in \cite
{Verlinde:2000wg}. Note that the total energy in cylindrical coordinates is
given by $E_{cyl}=E_{E}+E_{C}+E_{C,A}$. For $d=3$ one has $E_{C}=-E_{C,A}>0$
and it follows that $E_{cyl}=E_{E}$.

\subsection{Black holes in de Sitter}

We will now proceed to the case of de Sitter space to see whether similar
formulae are applicable to the relevant conjectured\ Euclidean CFT's.%
\footnote{%
For other recent attempts in this direction, see \cite
{Balasubramanian:2001nb}\cite{Myung:2001ab}\cite{Cai:2001jd}.} According to
our previous calculations the de Sitter black hole is a state with negative
energy. For convenience we define 
\[
E_{plan}=-\frac{d-2}{8\pi }\Omega _{d-2}M. 
\]
Using cylindrical coordinates, the new de Sitter vacuum has a shifted energy
due to the anomaly (if d is odd) given by 
\[
E_{cyl}=E_{C,A}-\frac{d-2}{8\pi }\Omega _{d-2}M. 
\]
As was explained above, the anomaly $E_{C,A}$ is positive and hence has the
opposite sign to AdS in the case of $d=3$, while the sign is the same in $%
d=5 $. If we now subtract the contribution from empty dS, just as in the AdS
case, we find 
\[
E_{E}+E_{C}=-\frac{d-2}{8\pi }\Omega _{d-2}M. 
\]
To be able to account for the entropy in de Sitter space we conclude,
following \cite{Verlinde:2000wg}, that the Casimir contribution is given by 
\[
E_{C}=-\frac{\left( d-2\right) c}{48\pi }\frac{V}{L^{d-3}R^{2}}, 
\]
which is \textit{negative} and therefore\textit{\ }has the \textit{opposite}
sign compared to the CFT relevant for AdS. This is also what to be expected
from the naive continuation $R^{2}\rightarrow -R^{2}$. The entropy is again
given by 
\[
S=\frac{4\pi R}{d-2}\sqrt{\left| E_{C}\right| \left( E_{plan}-E_{C}\right) }%
, 
\]
while the temperature becomes 
\begin{equation}
T=\frac{1}{4\pi L}\left( d-1-\left( d-3\right) \frac{L^{2}}{R^{2}}\right) ,
\label{dStemperature}
\end{equation}
where again $R^{2}$ has been replace by $-R^{2}$. This has important
consequences that we will come back to shortly. As in the case of AdS one
may note that $E_{C}=-E_{C,A}<0$ for $d=3$ implying $E_{cyl}=E_{E}$.

Let us now verify that the above indeed reproduces the correct result for
black holes. We will do the AdS and the de Sitter cases at the same time.
The prescription tells us to impose 
\[
\frac{\left( d-2\right) c}{48\pi }\frac{V}{L^{d-1}}\pm \frac{\left(
d-2\right) c}{48\pi }\frac{V}{R^{2}L^{d-3}}=\pm \frac{d-2}{8\pi }\Omega
_{d-2}M 
\]
where the + sign refers to AdS and the - sign refers to de Sitter. With $%
L=R^{2}/r_{s}$ (where $r_{s}$ is the horizon radius) and 
\[
c=3R^{d-2}, 
\]
the equation becomes 
\begin{equation}
1\pm \frac{r_{s}^{2}}{R^{2}}-\frac{2M}{r_{s}^{d-3}}=0,  \label{horizon}
\end{equation}
which is precisely the equation for the position of a horizon. From the
formula for the entropy we find 
\[
S=\frac{c}{12}\frac{V}{L^{d-2}}=\frac{1}{4}\Omega _{d-2}r_{s}^{d-2}=\frac{1}{%
4}A, 
\]
and hence the generalized Cardy-Verlinde formula can account for the black
hole entropy in AdS as well as de Sitter space. The case $d=3$ gives, using $%
E_{C,A}=\frac{c}{24R}$, 
\[
S=4\pi R\sqrt{E_{C,A}E_{E}}=2\pi \sqrt{\frac{c}{6}\left( RE_{plan}+\frac{c}{%
24}\right) }, 
\]
where $E_{plan}<0$. In the case of AdS one has $E_{plan}>0$ and the shift is
by -$\frac{c}{24}$.\footnote{%
Note that we are considering left and right movers together with $%
c=c_{L}+c_{R}$.}

There is a slight puzzle with the above construction. In the case of de
Sitter space there are two solutions of equation (\ref{horizon}), since
there are two horizons. One corresponds to the cosmological de Sitter
horizon, while the other corresponds to the horizon of the black hole. It is
easily checked, however, that it is only the cosmological horizon that leads
to a positive temperature. The conclusion, then, is that the CFT that we are
studying on $\mathcal{I}^{+}$ is accounting for the entropy of the
cosmological horizon only. This is consistent with the fact that the energy
and the entropy of the system is decreased when the mass of the black hole
is increased.\footnote{%
One may note that if the expression for the energy of the system (and
therefore also the temperature) is considered to have the opposite sign, the
reasoning is reversed. Positive temperature now suggests that it is the
entropy of the black hole horizon that is accounted for by the theory. This
is presumably the case for $\mathcal{I}^{-}$.} For more on the entropy of
black holes in de Sitter space see \cite{Bousso}. The total entropy of both
horizons as well as the entropy of the cosmological horizon have this
property, but our conclusion is that it is only the entropy of the
cosmological horizon that is accounted for by the Euclidean CFT on $\mathcal{%
I}^{+}$. Empty de Sitter space is also covered by the analysis and entails
the presence of a gas at nonzero temperature that can carry the entropy.

\subsubsection{The Nariai black hole}

The largest possible black hole in de Sitter space, the Nariai black hole 
\cite{Nariai}, corresponds to a situation where the two horizons coincide.
It is easy to check that this occurs when 
\[
M=\frac{1}{d-1}\left( \frac{d-3}{d-1}\right) ^{\frac{d-3}{2}}R^{d-3}, 
\]
which corresponds to zero temperature. Furthermore, on $\mathcal{I}^{+}$,
where the energy of a black hole is lower than that of empty de Sitter
space, the Nariai black hole corresponds to a lowest energy state. It would
be interesting to understand this better from the point of view of the
holographic theory. As observed in \cite{Klemm:2001ea} \cite
{Balasubramanian:2001nb} the energy in cylindrical coordinates vanishes for
the Nariai black hole in $d=5$, and a corresponding statement can also be
made in $d=3$. Is this just a coincidence or is there a deeper explanation?
This should be investigated further by comparing with higher dimensional
cases following, e.g., the analysis of \cite{Henningson:1998gx} and \cite
{Das:2000cu}. For other work in this context see e.g. \cite{Nojiri:2001mf} 
\cite{Nojiri:2001ae}.

\bigskip

\section{Conclusions}

\bigskip

In this paper we have found that a modified Cardy-Verlinde entropy formula
reproduces the entropy of the cosmological de Sitter horizon in the presence
of a black hole. But there are several unclear points. Why does the CFT on $%
\mathcal{I}^{+}$ only provide the entropy for the cosmological horizon? Is
it correct to discard the negative temperature solution? Is the role of $%
\mathcal{I}^{-}$ to take care of the black hole horizon?

Furthermore, the de Sitter horizon is an observer dependent construction.
How does the CFT take this into account? Our analysis gives a hint. Let us
consider a black hole at the origin of static coordinates. In planar
coordinates the energy density in the boundary theory corresponding to such
a black hole goes like $1/\rho ^{2}$.\footnote{%
Note that the subtracted Brown-York tensor is related to the energy-momentum
tensor of the CFT by an infinite conformal transformation. See, e.g., \cite
{Myers:1999qn}.} This is left invariant under a time translation in the bulk
in the same way as the metric. If we now take the point of view of an
observer who sees the black hole displaced from the center of de Sitter
space, the situation is different. In the case of a black hole with a
horizon much smaller than the cosmological horizon, and an observer far away
from the black hole, one can (for our purposes) approximate the system with
a black hole and an observer in free fall in de Sitter space. This means
that we can assume the distance between the two to be constant in comoving
coordinates. As time goes by the expansion will increase the proper distance
between the observer and the black hole. For the bulk observer time
translational invariance will be broken and she will see how the black hole
is approaching the cosmological horizon. As the black hole disappears from
view, the area of the cosmological horizon (according to our observer) will
increase with a net increase in entropy as a consequence. In the boundary
there is a similar story. The energy density is no longer left invariant
since the position of the maximum is offset from $\rho =0$ and will
effectively move off towards infinity as time goes by and we are zooming in
on smaller scales. In fact, for late times the relevant physics will be
taking place on small scales asymptotically independent of the black hole,
and energy as well as entropy will approach the values for empty de Sitter
space. It would be interesting to further study the time evolution of
entropy in this setting.

In \cite{Strominger:2001gp} it was indicated how a RG flow with a changing
central charge may take us from an inflationary era in the past with a small 
$R$ to a de Sitter space in the future with a large $R.$ But even with a
constant number of degrees of freedom it is interesting to study how non
trivial time evolution is encoded in various scales. What is the role of the
second law in the dS/CFT correspondence? How is the Hawking evaporation of
the black hole taken into account? There are clearly many interesting and
important questions that need to be investigated.

\section*{Acknowledgments}

I would like to thank Josef Kluson for discussions. The author is a Royal
Swedish Academy of Sciences Research Fellow supported by a grant from the
Knut and Alice Wallenberg Foundation. The work was also supported by the
Swedish Research Council (VR).

\end{document}